\documentclass[runningheads]{llncs}

\usepackage[T1]{fontenc}
\usepackage{cite}
\usepackage{amsmath,amssymb,amsfonts}
\usepackage{graphicx}
\usepackage[font=footnotesize]{subfig}
\usepackage{multirow}
\usepackage{url}
\usepackage{booktabs}
\usepackage{pifont}
\usepackage{xcolor}
\newcommand{\cmark}{\ding{51}}
\newcommand{\xmark}{\ding{55}}

\begin{document}

\title{Dynamic Adaptation of the LLM Context for Generating Routines with Coupled Semantics}

\titlerunning{Dynamic Adaptation of LLM Context for Coupled Semantics}

\author{Gnaneswar Villuri \and
Hashmath Shaik \and
Alex Doboli}
\authorrunning{G. Villuri et al.}
\institute{Department of ECE, Stony Brook University, Stony Brook, NY, USA\\
\email{gnaneswar.villuri@stonybrook.edu}}

\maketitle

\begin{abstract}
LLM-based code generation fails when correctness depends on execution-dependent coupling: the meaning of one routine is defined by the runtime behavior of another, a relationship that cannot be resolved from textual descriptions alone. This limitation, which we call \emph{static binding}, is not confined to explicitly coupled problems; it appears to varying degrees whenever correctness depends on joint execution behavior across components, from explicit cross-coupled optimizers to subtler joint constraints in packing, routing, and symbolic search. This paper proposes \emph{dynamic context adaptation}, a sample-efficient validation-generation loop designed for this setting. A validation agent extracts structured diagnostic information from execution traces, providing gradient-like guidance to a generation agent that proposes multiple candidates per iteration. A knowledge graph derived from the problem description supplies semantic constraints to the generation agent. Simulated annealing selects among candidates to avoid greedy collapse. Our method outperforms zero-shot, Reflexion, and OpenEvolve on seven of eight problems at both 300 and 600 evaluations ($p < 0.01$), a regime where population-based search has not yet accumulated sufficient diversity to compete. Notably, on the primary motivating problem (cross-coupled optimization), our method also achieves the best score at 1000 evaluations, consistent with the hypothesis that structured execution feedback is most beneficial when correctness depends on runtime coupling. Ablation results confirm that structured execution feedback is the primary driver.
\end{abstract}

\section{Introduction}
\label{introduction}

Automated software generation using LLMs has become a prominent research topic~\cite{dong2025survey,cui2024effects,chen2025codewithme}, with tools, e.g., Cursor~\cite{krishnaswami2025cursor,jiang2024cursorcore}, offering substantial productivity gains~\cite{peng2023copilot,pandey2024copilot}. LLM-based code generation excels at scaffolding but misses execution-dependent patterns~\cite{jimenez2024}. Evolutionary methods (like genetic programming~\cite{koza1994}, FunSearch~\cite{funsearch2023}, AlphaEvolve~\cite{alphaevolve}) use mutation operators but require thousands of evaluations. Feedback-driven methods, i.e. CEGIS~\cite{solar2008} and Reflexion~\cite{shinn2023}, iterate from execution outcomes, but Reflexion operates on a \emph{single} trajectory without structured diagnostic extraction or multi-candidate generation, and commits greedily to improvements. Moreover, a persistent limitation remains: LLMs struggle when the correctness of one component depends on the \emph{runtime behavior} of another, not just its textual description. We call this \emph{static binding}: 
inter-component meaning dependencies can be resolved only by observing the execution patterns of the components.

The code generation method proposed in this paper uses a validation-generation agent loop with a knowledge-graph (KG) intermediate layer and simulated annealing (SA) for candidate selection to generate multiple candidates per iteration, enabling broader exploration (see Section~\ref{system_architecture}). To address the limitations of static binding, a dedicated validation agent extracts \emph{structured} feedback from code execution traces rather than from free-form verbal reflection. SA balances exploration and exploitation instead of making greedy commitments, as in most current approaches. The central claim is \emph{sample efficiency}: structured execution feedback produces rapid, directed improvement before population-based diversity can accumulate. Our convergence curves locate the crossover with population-based search empirically at 600--800 evaluations; below that threshold our system outperforms all iterative baselines on seven of eight problems at both 300 and 600 evaluations ($p < 0.01$, Table~\ref{tab:results_summary}, Figure~\ref{fig:convergence}). The crossover reflects a complementary tradeoff between feedback-driven early convergence and population-based asymptotic exploration, not a failure of the feedback mechanism. 

This paper makes the following contributions:
\begin{enumerate}
\vspace * {-0.1in}
\item It presents a formal taxonomy of meaning dependencies among software components, providing a precise characterization of the LLM static-binding limitation. Four related cases are discussed in Section~\ref{problem_description}.
\item It proposes a novel two-agent, KG-based system architecture for code synthesis (Section~4).
\item It provides empirical evidence of superior sample efficiency: our method outperforms all iterative baselines on seven of eight problems at both 300 and 600 evaluations ($p < 0.01$), with the feedback-vs-diversity crossover located empirically at 600--800 evaluations (Section~\ref{experiments}).
\end{enumerate}

\begin{table}[!t]
\caption{Qualitative comparison of LLM-based code generation approaches.}
\label{tab:method_comparison}
\centering
\small
\setlength{\tabcolsep}{6pt}
\begin{tabular}{lcccc}
\toprule
\textbf{Property} & \textbf{Zero-shot} & \textbf{Reflexion}~\cite{shinn2023} & \textbf{AlphaEvolve}~\cite{alphaevolve} & \textbf{Ours} \\
\midrule
Iterative refinement      & \xmark & \cmark & \cmark & \cmark \\
Structured feedback       & \xmark & \xmark & \xmark & \cmark \\
Multi-candidate per iter. & \xmark & \xmark & \cmark & \cmark \\
Non-greedy selection      & \xmark & \xmark & \cmark & \cmark \\
Semantic layer (KG)       & \xmark & \xmark & \xmark & \cmark \\
\bottomrule
\end{tabular}
\end{table}

\vspace * {-0.2in}
\section{Motivation}
\label{motivation}


The next two examples 
illustrate the same limitation: LLMs link concepts at generation time without interpreting their runtime semantics. Conceptually, this resembles static parameter binding, e.g., formal parameters are bound to actuals at compile time with no interpretation of runtime values. LLM attention is similarly fixed during training. However, when correctness depends on execution patterns (e.g., the robot actions in Example~1 and the coupled optimizers in Example~2), meaning must be discovered dynamically. 

{\bf Example 1: Maze navigation}:
\label{maze_problem}
The problem requires devising a maze-navigation algorithm so that an autonomous robot collects the maximum number of items. An item can be collected only after its specific key is first collected, and thus keys must be reached before the items. The robot can execute seventeen specific actions, like moving one position to the left, right, up, and down if there is no obstacle (e.g., a wall), marking a place and storing its position in the memory, checking if a place has a marking, aggressively jumping over a number of spots until reaching an obstacle, retrieving a stored position from the memory, and loop instructions. 

A frontier commercial LLM was used to solve the problem. It correctly identified the main components of the design, like the scanner routines to read the maze file and the navigation method, and the routine that implements the seventeen instructions.
Prompting the LLM to explain its reasoning indicated that it used a backtracking algorithm, which attempts the four possible moves at each step (move left, right, up, and down), stores the current position of the robot in the memory, and backtracks if an impasse is reached, e.g., navigation cannot continue. However, the following issues were observed: (i)~navigation in terms of the robot actions did not match the conceptual backtracking algorithm, (ii)~the backtracking strategy was random, e.g., the robot backtracks after every 200 moves independently of the effectiveness of its search, and (iii)~only seven of the robot actions were used, although the navigation could have benefited from other actions as well.  

{\bf Example 2: Cross-coupled optimization}.
\label{multi_optimization_problem}
The second problem concerned resource allocation for managing disaster situations, e.g., a neighborhood power-grid outage. The goal was to optimize the number of patients treated by the neighborhood hospitals considering that each hospital has a number of surgical teams with different levels of expertise and efficiency and various energy storage options. In addition, the optimization goal had to be correlated with the way in which the companies that service the power grid repair the damages. The optimization goal (number of treated patients) is strongly coupled with two other optimization needs, i.e., the energy resources assigned to a surgical team (depending on its expertise and efficiency) and the sequence of addressing the damages depending on the available crews.

The LLM could correctly generate code for all data management routines, including the description of grid damages, surgery teams, and hospital energy storage options. However, the four optimization algorithms were handled independently as the LLM did not understand that solving the problem requires coupling the four optimization methods, so that the decisions of each method are linked to the decisions of the other procedures. 

\begin{figure}[!t]
  \centerline{\includegraphics[width=0.60\textwidth]{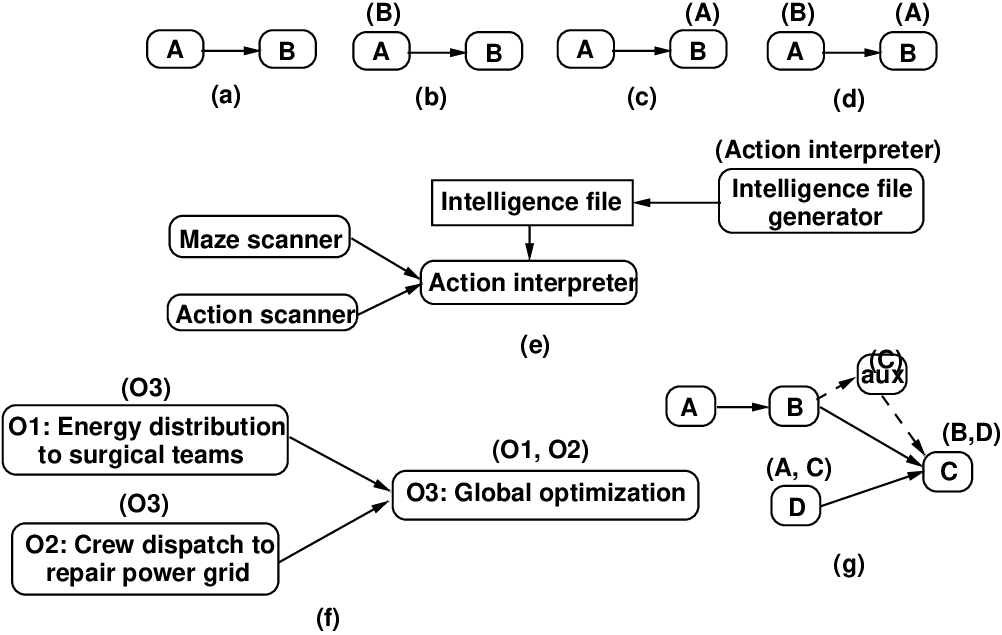}}
  \caption{Meaning dependencies among components: (a)-(d) the four cases; (e) Example~1; (f) Example~2; (g) general case with auxiliary $aux$.}
  \label{fig1}
  \vspace * {-0.1in}
  \end{figure}
  \vspace * {-0.1in}
  \section{Problem Description}
  \label{problem_description}
  
This work describes {\em problem understanding} as finding the LLM context that supports the correct and efficient solving of the problem. Problem understanding includes interpreting functional requirements, performance requirements (e.g., execution time, memory, power/energy consumption), constraints on code organization and data structures, required testing and edge cases, and future needs to maintain and modify the code \cite{pressman2014}. The LLM contexts are found through prompting strategies, e.g. through adaptive sequences of prompts.    

The design and implementation of the system architecture involves the identification of the related modules (components) and their functions, connecting the modules with each other, the identification of the algorithmic approaches for each component, the decomposition of complex functions into simpler steps, and finally the implementation of the simpler steps \cite{sommerville2016}.

Finding the correct context for problem understanding and solving requires identifying the context layers that address the LLM limitations summarized in Section~2, including the static linking of the concepts and relations. The meaning (semantics) of the components of the architecture must be interpreted in the correct context, which sometimes involves the meaning of other components too, as was discussed for Example~1 (maze navigation). The meaning cannot always be reasoned out statically from the problem description. Instead, it can only be obtained after executing the component and observing its behavior. 

Figure~\ref{fig1} summarizes the four situations that occur when deciding the meaning of a component. They decide the procedure to identify the correct context:
\begin{enumerate}
\item 
{\em The independent case}: The meaning of a component is fully defined by its description and does not depend on other components. Figure~\ref{fig1}(a) shows the component~$B$ providing inputs to component~$A$, however, the meaning of the two routines do not depend on each other. Their only connection is through the data they share. This is the case of the two scanner routines in Example~1, as their meaning does not depend on another routine, and hence are independent. The next equation gives a formal description: 
\begin{equation}
\mathcal{M} (A), \mathcal{M} (B) \nparallel \mathcal{M} (R), \forall R
\label{eq1}
\end{equation}
The equation states that the meaning ($\mathcal{M}$) of components $A$ and $B$ are independent of the meaning of any routine~$R$ of the description or solution. 

\item 
{\em Sender related to the receiver}: As shown in Figure~\ref{fig1}(b), the meaning of the component~$A$ (data sender/originator) depends on the semantics of the data receiver component~$B$. This is the case of the maze navigation routine in Example~1, as the meaning of the generator, and thus its actions, depends on the robot actions. The next equation offers a formal description: 
\begin{equation}
\mathcal{M} (A) \propto \mathcal{M}(B); \mathcal{M} (B) \nparallel \mathcal{M} (R), \forall R
\label{eq2}
\end{equation}
The meaning of component~$A$ is related ($\propto$) to the meaning of component $B$, but the meaning of $B$ is independent of the meaning of any routine~$R$. 

\item 
{\em Receiver related to the sender}: Figure~\ref{fig1}(c) depicts this case. The situation is similar to the previous case, but now the meaning of the data receiving component depends on the meaning of the data generating component. The next equation gives a formal description: 
\begin{equation}
\mathcal{M} (B) \propto \mathcal{M}(A); \mathcal{M} (A) \nparallel \mathcal{M} (R), \forall R
\label{eq3}
\end{equation}

\item 
{\em Cross-correlated sender and receiver}: As shown in Figure~\ref{fig1}(d), the meanings of the components~$A$ and~$B$ depend on each other. Hence:
\begin{equation}
\mathcal{M} (B) \propto \mathcal{M}(A); \mathcal{M} (A) \propto \mathcal{M} (B)
\label{eq4}
\end{equation}
\end{enumerate}
Figure~\ref{fig1}(e) presents the meaning dependencies for Example~1 (maze navigator), and Figure~\ref{fig1}(f) shows the meaning dependencies of the three optimization routines in Example~2 (cross-coupled optimization). Finally, Figure~\ref{fig1}(g) illustrates the general case for defining the meaning of routines. In addition to the four cases, the implementation might require other components, like the routine~$aux$, not mentioned in the description, but needed to bridge between components. 

Static linking of concepts (e.g., routines, variables, and actions) by an LLM implicitly assumes that the meanings of the concepts are independent (equation~(\ref{eq1})). However, it is not always the case. Assuming that $\mathcal{M^*}(A)$ and $\mathcal{M^*}(B)$ are independent meanings of routines $A$ and $B$ produced by an LLM, the goals of context finding are to transform (i)~$\mathcal{M^*}(A)$ into $\mathcal{M}(A)$, such that $\mathcal{M} (A) \propto \mathcal{M}(B)$ for the second case, (ii)~$\mathcal{M^*}(B)$ into $\mathcal{M}(B)$, such that $\mathcal{M} (B) \propto \mathcal{M}(A)$ for the third case, and (iii)~$\mathcal{M^*}(A)$ into $\mathcal{M}(A)$ and $\mathcal{M^*}(B)$ into $\mathcal{M}(B)$, such that $\mathcal{M} (A) \propto \mathcal{M}(B)$ and $\mathcal{M} (B) \propto \mathcal{M}(A)$ for the fourth case.

The system architecture discussed in Section~4 starts with routine interpretations and implementations that are considered to be independent. Then, routine transformations corresponding to cases 2-4 are handled by a loop that uses a validating agent and test vectors to convert the independent meaning~$\mathcal{M^*}(R)$ of a routine $R$ into its equivalent meaning $\mathcal{M}(R) \propto \mathcal{M}(S)$, which relates to the meaning of routine~$S$. 

\section{System Architecture}
\label{system_architecture}

\begin{figure}[!t]
\centerline{\includegraphics[width=0.80\textwidth]{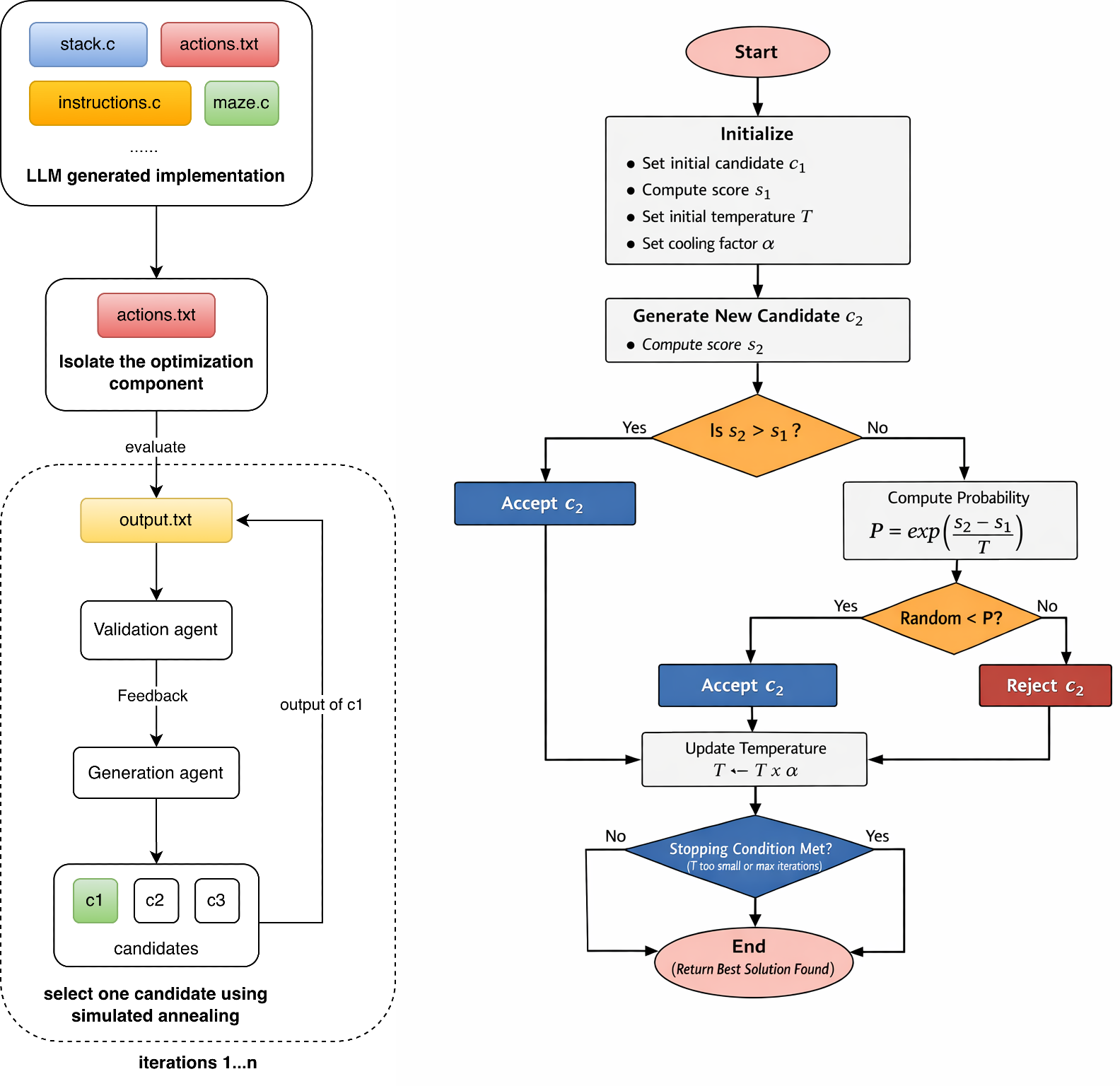}}
\caption{Left: validation-generation pipeline. Right: SA selection with Metropolis acceptance and geometric cooling.}
\label{fig3}
\end{figure}

Figure~\ref{fig3} shows the overall pipeline. Starting from an LLM-generated codebase (e.g., \texttt{stack.c}, \texttt{instructions.c}, \texttt{maze.c}), we isolate the optimization-sensitive component (e.g., \texttt{actions.txt}) and place it inside an iterative refinement loop. At each iteration, a \textbf{validation agent} executes the current candidate to produce output logs, analyzes quantitative outcomes and qualitative execution trace patterns, and produces structured diagnostic feedback summarizing failure modes, missed opportunities, and constraint violations. A \textbf{generation agent} receives this feedback alongside the current incumbent candidate and KG-derived constraints, and proposes $k$ new candidates ($c_1, c_2, \ldots, c_k$). Each generation prompt includes: (i)~a task summary from the problem description, (ii)~the KG-derived action semantics and coupling constraints, (iii)~the current incumbent, (iv)~execution traces and output logs, and (v)~validation feedback distilled into actionable suggestions. An SA controller then selects one candidate to carry forward, and the loop repeats until a stopping criterion is reached.

\textbf{Knowledge graph (KG).} To make meaning dependencies explicit, we extract a KG from the problem description via a two-step LLM prompt: first clustering text into structure, requirements, constraints, and artifacts; then extracting typed nodes $(\text{id}, \text{label}, \text{type}, \text{attributes})$ and typed edges (implements, constrains, depends\_on, etc.) with hard/soft strength tags. The relevant subgraph is selected by dependency tracing from the optimization-sensitive node and inserted as structured JSON into each agent prompt, yielding $\mathcal{M}_\text{text} \rightarrow \mathcal{M}_\text{KG} \rightarrow \mathcal{M}_\text{code}$.

\begin{figure}[!t]
\centering
\subfloat[Maze navigation KG.]{%
\includegraphics[width=0.45\textwidth]{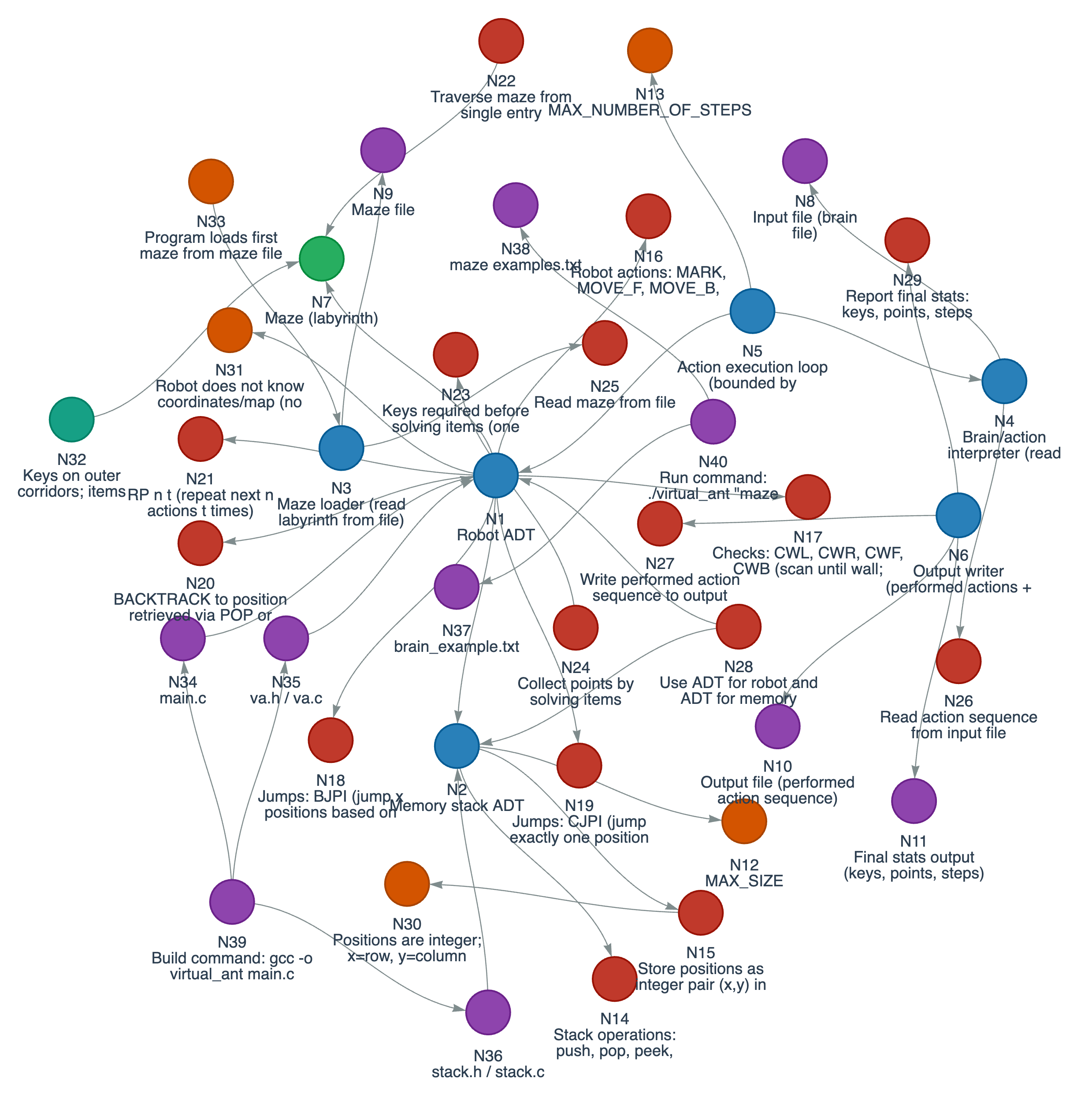}}
\hfill
\subfloat[Cross-coupled optimization KG.]{%
\includegraphics[width=0.45\textwidth]{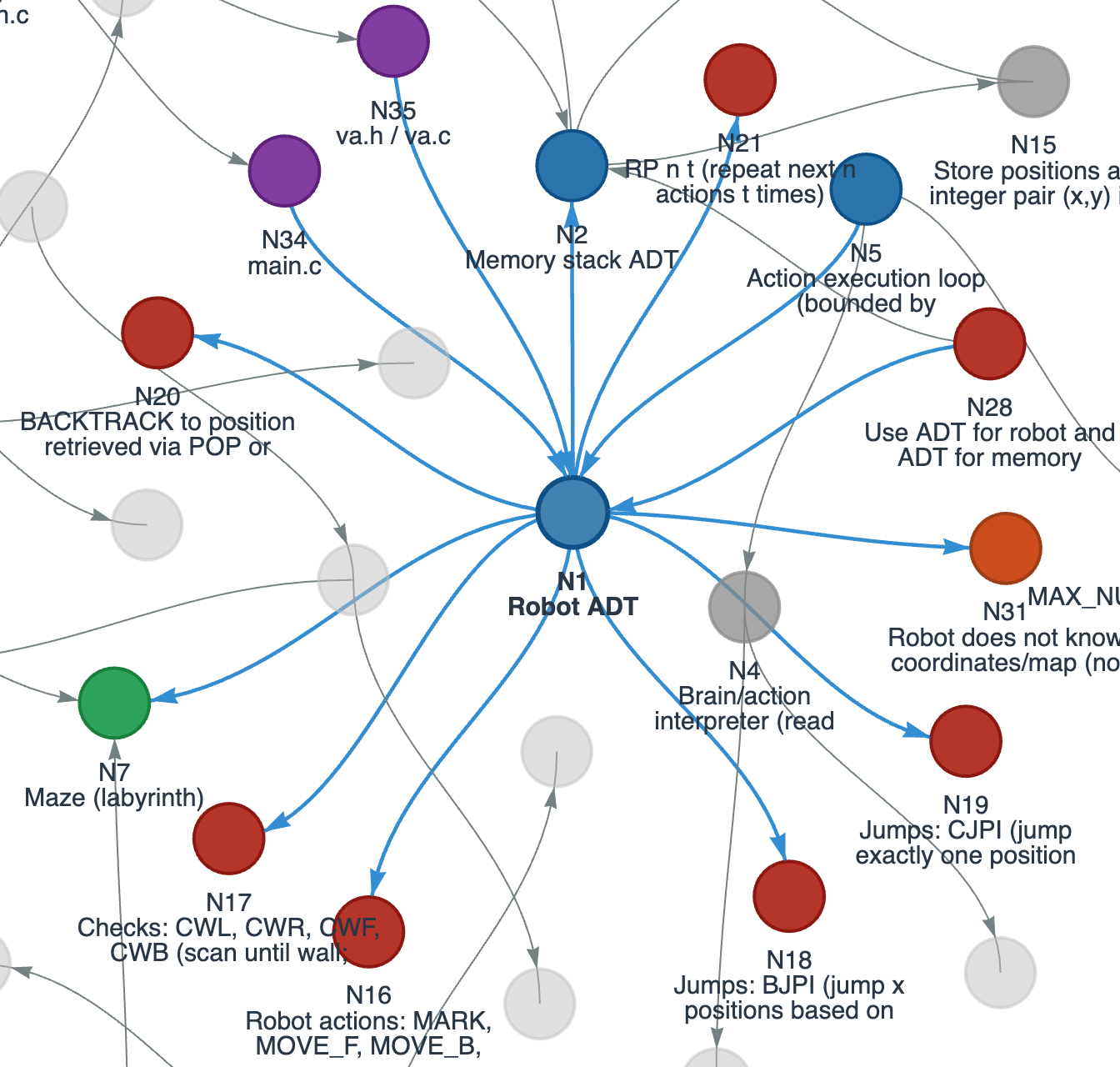}}
\caption{Knowledge graphs showing modules, requirements, and meaning dependencies.}
\label{fig:kg_examples}
\end{figure}

\textbf{SA selection.} Let $s_1$ be the current incumbent score; a new candidate with score $s_2$ is accepted unconditionally if $s_2 > s_1$, otherwise with probability $P = \exp((s_2-s_1)/T)$ compared to Uniform$(0,1)$. Temperature cools geometrically: $T \leftarrow T \times \alpha$, with $T_0=2.5$, $\alpha=0.85$, $T_{\min}=0.01$ (matching the experimental parameters in Section~\ref{experiments}). SA prevents collapse into locally adequate policies and allows bold structural changes that greedy selection cannot make. The cooling rate $\alpha=0.85$ front-loads exploration in the first $\sim$30 iterations before structured feedback has accumulated sufficient signal; beyond that point, near-greedy selection is appropriate since the gradient-like feedback increasingly constrains the viable candidate space.

\section{Problems and Cost Functions}
\label{implementation}

We evaluate on eight problems. \textbf{P1} and \textbf{P2} are original problems introduced in Section~\ref{motivation}, where coupled-semantics dependencies make LLM static binding a concrete, measurable limitation. \textbf{P3--P8} are standard benchmarks included for direct comparison with prior work; their formulations follow the cited sources without modification.

\textbf{P1: Maze navigation} (Case~2 dependency). Three maze instances each award up to 10~points under key-precedence constraints, giving a theoretical maximum of 30~points total. The generation agent is constrained strictly to the 17-instruction ISA (\texttt{MOVE\_*}, \texttt{MARK}, \texttt{CWL/R/F/B}, \texttt{PUSH/POP/PEEK/CLEAR}, \texttt{BJPI}, \texttt{CJPI}, \texttt{BACKTRACK}, \texttt{RP~n~t}). The validation agent inspects per-maze execution logs for unproductive loops, wasted stack operations, and invalid token usage, distilling findings into actionable suggestions.
\begin{equation}
\mathcal{C}_{\text{maze}} = \textstyle\sum_{i=1}^{3} \text{score}_i, \quad \text{score}_i \in [0,10],\quad \mathcal{C}_{\text{maze}} \in [0,30]
\label{eq_maze_cost}
\end{equation}

\textbf{P2: Cross-coupled optimization} (Case~4 dependency). Joint optimization of power distribution to surgical teams and crew dispatch for grid repair. Weights $w_1{=}0.5, w_2{=}0.3, w_3{=}0.2$ reflect clinical priority; $\mathcal{C}_{\text{couple}}{=}1.0$ corresponds to baseline performance. The validation agent flags coupling violations (e.g., surgeries during active outages) and passes structured reports to the generation agent.
\begin{equation}
\mathcal{C}_{\text{couple}} = w_1 \tfrac{\Phi}{\Phi_{\text{ref}}} + w_2 \tfrac{H_{\text{ref}}}{H} + w_3 \tfrac{\Pi}{\Pi_{\text{ref}}}
\label{eq_couple_obj}
\end{equation}

\textbf{P3--P8: Standard benchmarks.} Circle packing and function minimization are taken from the OpenEvolve benchmark suite~\cite{openevolve,alphaevolve}. TSP tour minimization~\cite{helsgott2000}, signal processing filter design~\cite{oppenheim1999}, online judge programming~\cite{jimenez2024}, and symbolic regression~\cite{koza1994} follow standard formulations. All six use a normalized score in $[0,1]$; cost functions maximize solution quality relative to the known optimum or a reference solution. Although not explicitly framed as coupled-semantics tasks, each exhibits latent execution-dependent coupling (e.g., circle packing radii are jointly constrained; TSP tour cost depends on the full sequence), which is why structured execution feedback yields improvements even on these standard benchmarks.

\section{Experiments}
\label{experiments}

\subsection{Setup}

All experiments use Qwen/Qwen2.5-72B-Instruct served locally (temperature\,=\,1.0, top-$p$\,=\,1.0, max 8{,}192 tokens). This model was chosen because its training data predates the release of OpenEvolve, eliminating any risk of benchmark contamination from the OpenEvolve problem suite. We use SA parameters $T_0{=}2.5$, $\alpha{=}0.85$, and $T_{\min}{=}0.01$. Runs extend to 333 iterations $\times$ 3 candidates $\approx$ 1000 evaluations per problem. Running a locally served open-weight model makes this tractable without API cost: seven of eight problems complete in under 20 minutes (300 evaluations $\approx$ 5--8 minutes), fitting within a developer iteration cycle; circle packing is the outlier at $\sim$94 minutes due to a 600\,s sandbox cap. Table~\ref{tab:results_summary} reports all methods at 1000 evaluations; Table~\ref{tab:convergence} compares OpenEvolve and ours at 300, 600, and 1000 evaluations; Figure~\ref{fig:convergence} shows the full trajectory. Each condition uses 10 random seeds; we report mean\,$\pm$\,std. We use a paired Wilcoxon signed-rank test across seeds to assess statistical significance and report $p$-values alongside effect sizes.

\textbf{Baselines:}
\begin{itemize}
  \item \textbf{Zero-shot}: single direct prompt, no loop (budget-independent).
  \item \textbf{Reflexion}~\cite{shinn2023}: single-trajectory verbal RL, greedy acceptance, no KG.
  \item \textbf{OpenEvolve}~\cite{openevolve}: open-source implementation of AlphaEvolve~\cite{alphaevolve}.
\end{itemize}
AlphaEvolve~\cite{alphaevolve} ($\approx$1.0004 on circle packing) is included as a quality ceiling only; it uses proprietary Gemini models with substantially greater compute and is not a controlled baseline.

\subsection{Main Results}

\begin{table}[!t]
\caption{Results at 1000 evaluations (mean\,$\pm$\,std, 10 seeds, Qwen2.5-72B). Bold: best per problem. Dep.: primary meaning-dependency case (§\ref{problem_description}); P1 and P2 have explicit coupled semantics (cases~2 and~4), while P3--P8 exhibit latent execution-dependent coupling (case~1 in primary structure).}
\label{tab:results_summary}
\centering
\footnotesize
\setlength{\tabcolsep}{2pt}
\begin{tabular}{|l|l|c|c|c|c|c|}
\hline
\textbf{Problem} & \textbf{Type} & \textbf{Dep.} & \textbf{Zero-shot} & \textbf{Reflexion} & \textbf{OpenEvolve} & \textbf{Ours} \\
\hline
Maze Nav.      & Discrete plan.   & 2 & 11.7  & 23.1\tiny$\pm$2.3  & \textbf{28.3\tiny$\pm$1.8}  & 27.1\tiny$\pm$1.2  \\
Circle Pack.   & Comb.\ geo.      & 1 & 0.364  & 0.869\tiny$\pm$0.051 & \textbf{0.978\tiny$\pm$0.011} & 0.961\tiny$\pm$0.018 \\
Func.\ Min.    & Cont.\ opt.      & 1 & 1.215 & 1.481\tiny$\pm$0.031 & \textbf{1.519\tiny$\pm$0.008} & 1.514\tiny$\pm$0.009 \\
Cross-Coupled  & Multi-obj.\ opt. & 4 & 0.511  & 0.603\tiny$\pm$0.027 & 0.681\tiny$\pm$0.028 & \textbf{0.694\tiny$\pm$0.019} \\
TSP            & Comb.\ opt.      & 1 & 0.398  & 0.644\tiny$\pm$0.046 & \textbf{0.803\tiny$\pm$0.024} & 0.775\tiny$\pm$0.027 \\
Signal Proc.   & Filter design    & 1 & 0.291  & 0.528\tiny$\pm$0.054 & \textbf{0.715\tiny$\pm$0.039} & 0.692\tiny$\pm$0.038 \\
Online Judge   & Prog.\ synth.    & 1 & 0.238  & 0.496\tiny$\pm$0.058 & \textbf{0.702\tiny$\pm$0.041} & 0.679\tiny$\pm$0.031 \\
Symbolic Reg.  & Func.\ disc.     & 1 & 0.324  & 0.584\tiny$\pm$0.050 & \textbf{0.781\tiny$\pm$0.032} & 0.757\tiny$\pm$0.024 \\
\hline
\end{tabular}
\end{table}

\begin{table}[!t]
\caption{Convergence comparison: OpenEvolve (OE) vs.\ Ours at 300, 600, and 1000 evaluations (mean\,$\pm$\,std, 10 seeds). Bold: best per problem at each budget.}
\label{tab:convergence}
\centering
\footnotesize
\setlength{\tabcolsep}{1.8pt}
\begin{tabular}{|l|cc|cc|cc|}
\hline
 & \multicolumn{2}{c|}{\textbf{300 evals}} & \multicolumn{2}{c|}{\textbf{600 evals}} & \multicolumn{2}{c|}{\textbf{1000 evals}} \\
\textbf{Problem} & \textbf{OE} & \textbf{Ours} & \textbf{OE} & \textbf{Ours} & \textbf{OE} & \textbf{Ours} \\
\hline
Maze Nav.      & 19.2\tiny$\pm$3.1 & \textbf{25.8\tiny$\pm$1.4} & 23.3\tiny$\pm$2.5 & \textbf{26.8\tiny$\pm$1.3} & \textbf{28.3\tiny$\pm$1.8} & 27.1\tiny$\pm$1.2 \\
Circle Pack.   & 0.903\tiny$\pm$0.035 & \textbf{0.952\tiny$\pm$0.024} & 0.937\tiny$\pm$0.025 & \textbf{0.959\tiny$\pm$0.020} & \textbf{0.978\tiny$\pm$0.011} & 0.961\tiny$\pm$0.018 \\
Func.\ Min.    & \textbf{1.519\tiny$\pm$0.014} & 1.511\tiny$\pm$0.011 & \textbf{1.519\tiny$\pm$0.011} & 1.513\tiny$\pm$0.010 & \textbf{1.519\tiny$\pm$0.008} & 1.514\tiny$\pm$0.009 \\
Cross-Coupled  & 0.563\tiny$\pm$0.042 & \textbf{0.654\tiny$\pm$0.021} & 0.621\tiny$\pm$0.035 & \textbf{0.665\tiny$\pm$0.020} & 0.681\tiny$\pm$0.028 & \textbf{0.694\tiny$\pm$0.019} \\
TSP            & 0.701\tiny$\pm$0.036 & \textbf{0.761\tiny$\pm$0.033} & 0.747\tiny$\pm$0.030 & \textbf{0.772\tiny$\pm$0.029} & \textbf{0.803\tiny$\pm$0.024} & 0.775\tiny$\pm$0.027 \\
Signal Proc.   & 0.541\tiny$\pm$0.055 & \textbf{0.678\tiny$\pm$0.044} & 0.619\tiny$\pm$0.047 & \textbf{0.689\tiny$\pm$0.041} & \textbf{0.715\tiny$\pm$0.039} & 0.692\tiny$\pm$0.038 \\
Online Judge   & 0.478\tiny$\pm$0.058 & \textbf{0.664\tiny$\pm$0.037} & 0.579\tiny$\pm$0.049 & \textbf{0.675\tiny$\pm$0.034} & \textbf{0.702\tiny$\pm$0.041} & 0.679\tiny$\pm$0.031 \\
Symbolic Reg.  & 0.639\tiny$\pm$0.049 & \textbf{0.743\tiny$\pm$0.028} & 0.703\tiny$\pm$0.041 & \textbf{0.754\tiny$\pm$0.025} & \textbf{0.781\tiny$\pm$0.032} & 0.757\tiny$\pm$0.024 \\
\hline
\end{tabular}
\end{table}

\begin{figure}[!t]
\centerline{\includegraphics[width=\columnwidth]{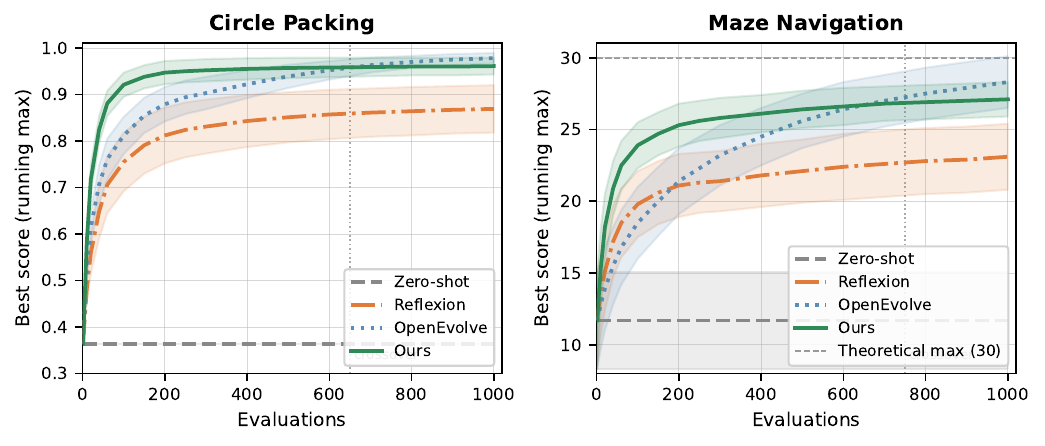}}
\caption{Running maximum (best-so-far) score vs.\ evaluation count (mean $\pm$1 std, 10 seeds) for circle packing (left) and maze navigation (right). Our method outperforms OpenEvolve up to $\sim$650 evaluations (circle packing) and $\sim$710 evaluations (maze navigation); OpenEvolve surpasses it at 1000 evaluations.}
\label{fig:convergence}
\end{figure}

Table~\ref{tab:results_summary} shows the full-budget comparison. At 1000 evaluations OpenEvolve outperforms our method on seven of eight problems (significant on six, $p < 0.05$; function minimization: $p = 0.091$, $d = 0.29$); the exception is cross-coupled optimization (P2), where our method achieves the highest score ($0.694$ vs.\ $0.681$, $p = 0.019$, $d = 0.52$), consistent with the hypothesis that structured execution feedback is most effective precisely when correctness depends on runtime coupling. Our method places second on the remaining seven, significantly outperforming Reflexion on all eight ($p < 0.001$ on six; $p \leq 0.007$ on the remaining two). Table~\ref{tab:convergence} reveals the convergence dynamics: at 300 evaluations our method outperforms OpenEvolve on seven of eight problems (all $p < 0.01$; function minimization is the exception, $p = 0.284$, where OpenEvolve converges within the first 100 evaluations), and this advantage holds at 600 evaluations on the same seven problems (all $p < 0.01$). The convergence curves in Figure~\ref{fig:convergence} show the underlying mechanism: structured feedback drives steep early improvement that plateaus once the incumbent stabilizes, while OpenEvolve's population requires a burn-in period but finds novel structural variations beyond the crossover (${\sim}650$ evaluations for circle packing, ${\sim}710$ for maze). The effect is most pronounced on multi-basin problems and least on smooth landscapes (function minimization). Reflexion plateaus well below both methods ($p < 0.001$) due to its greedy single-trajectory search.

\subsection{Ablation Study}

\begin{table}[!t]
\caption{Ablation study at 300 evaluations (mean\,$\pm$\,std, 10 seeds). Full system in bold.}
\label{tab:ablation}
\centering
\small
\setlength{\tabcolsep}{4pt}
\begin{tabular}{|l|c|c|c|c|}
\hline
\textbf{Condition} & \textbf{Maze} & \textbf{Circle} & \textbf{Func Min.} & \textbf{Cross-Cpl.} \\
\hline
\textbf{Full (Ours)} & \textbf{25.8$\pm$1.4} & \textbf{0.952$\pm$0.024} & \textbf{1.511$\pm$0.011} & \textbf{0.654$\pm$0.021} \\
w/o Feedback          & 18.6$\pm$2.8          & 0.657$\pm$0.081          & 1.330$\pm$0.058          & 0.582$\pm$0.029          \\
Greedy (no SA)        & 21.7$\pm$2.9          & 0.866$\pm$0.097          & 1.481$\pm$0.038          & 0.615$\pm$0.037          \\
Random selection      & 14.4$\pm$4.1          & 0.533$\pm$0.152          & 1.209$\pm$0.131          & 0.536$\pm$0.051          \\
w/o KG                & 21.9$\pm$2.1          & 0.895$\pm$0.058          & 1.481$\pm$0.029          & 0.628$\pm$0.033          \\
\hline
\end{tabular}
\end{table}

\textbf{Structured feedback is the largest single contributor.} Removing it causes the largest drop across all four problems ($-28\%$ maze, $-31\%$ circle, $-12\%$ function minimization, $-11\%$ cross-coupled; all $p < 0.001$, $d > 1.1$). Without diagnostic feedback the generation agent degrades to near-random candidate generation; the ``w/o Feedback'' row approaches the ``Random selection'' lower bound. \textbf{SA selection} matters most on high-variance discrete problems: replacing SA with greedy selection reduces maze navigation by $-16\%$ ($p < 0.001$, $d = 0.89$) and circle packing by $-9\%$ ($p = 0.002$, $d = 0.71$), while increasing standard deviation across all conditions, reflecting premature convergence. On the smoother function minimization landscape the SA effect is smaller but still significant ($p = 0.018$, $d = 0.44$). \textbf{The KG} provides a consistent but moderate gain ($-15\%$ maze, $p = 0.001$, $d = 0.78$; $-6\%$ circle, $p = 0.014$, $d = 0.51$), most beneficial when rich constraint structure would otherwise be underspecified in plain text.

\section{Discussion}
\label{discussion}

\subsection{Structured Feedback as Gradient Approximation}

The validation agent's structured output (e.g., ``overlaps between circles 1 and 12,'' ``MOVE\_D token is not in the ISA and was ignored'') provides directional guidance that is qualitatively analogous to gradients in continuous optimization: it identifies both the location and the cause of failure. A recurring pattern was a transition from local, per-component edits to globally coupled solutions: in circle packing, the breakthrough was treating all 26 circles as a jointly constrained system, a relationship only observable at execution time, precisely the static-binding failure our loop targets. This allows the generation agent to make focused, targeted corrections rather than undirected mutations. This gradient-like property explains why structured feedback dominates in the low-budget regime: rather than relying on population diversity to explore the search space, the generation agent receives focused directional signals that eliminate entire classes of failure before the next iteration, compressing what would otherwise require hundreds of undirected mutations into a handful of targeted corrections.

\subsection{Limitations}

The validation agent diagnoses surface-level symptoms without counterfactual reasoning; discrete action spaces exhibit high variance because single-token changes produce large behavioral shifts. Feedback scope is bounded by the test suite, so latent bugs outside the execution log escape detection. The single-incumbent design limits diversity on multi-basin landscapes, where population-based methods have a structural advantage.

\section{Conclusion}

We proposed \emph{dynamic context adaptation}: a validation-generation loop that grounds LLM code generation in execution feedback rather than static linking. The framework introduces a four-case taxonomy of meaning dependencies, a KG intermediate layer, and SA selection across multiple candidates per iteration. It outperforms all iterative baselines on seven of eight problems at both 300 and 600 evaluations, and achieves the best full-budget score on cross-coupled optimization (the primary motivating problem), where runtime coupling most directly favors structured feedback. The crossover with population-based search at 600--800 evaluations reveals a complementary tradeoff. Ablation results confirm structured execution feedback as the primary driver, followed by SA selection and KG grounding.

Future work includes extending the taxonomy to multi-hop dependency chains, maintaining a small population of incumbents to recover diversity on multi-basin landscapes, and combining structured trace feedback with formal verification oracles.

\end{document}